\documentclass[conference]{IEEEtran}
\IEEEoverridecommandlockouts
\usepackage[utf8]{inputenc}
\usepackage{cite}
\usepackage{amsmath,amssymb,amsfonts}
\usepackage{algorithmic}
\usepackage{graphicx}
\usepackage{textcomp}
\usepackage{xcolor}
\usepackage{tabularx}
\usepackage{array} 
\newcolumntype{L}[1]{>{\raggedright\arraybackslash}p{#1}}
\usepackage{pgfplots}
\usepackage{pgfplotstable}
\pgfplotsset{compat=1.18}
\usepackage{sansmath}
\usepackage{adjustbox}
\usepackage{multirow}
\usepackage{booktabs}
\usepackage{float}
\usepackage[acronym, nopostdot]{glossaries}
\glsdisablehyper
\newacronym{FMA}{FMA}{Fused Multiply Accumulate}
\newacronym{SIMD}{SIMD}{Single Instruction Multiple Data}
\newacronym{FPGA}{FPGA}{Field Programmable Gate Array}
\newacronym{ASIC}{ASIC}{Application Specific Integrated Circuit}
\newacronym{CPU}{CPU}{Central Processing Unit}
\newacronym{GPU}{GPU}{Graphics Processing Unit}
\newacronym{HLS}{HLS}{High Level Synthesis}
\newacronym{RTL}{RTL}{Register Transfer Level}
\newacronym{VHDL}{VHDL}{VHSIC Hardware Description Language}
\newacronym{BCPNN}{BCPNN}{Bayesian Confidence Propagation Neural Network}
\newacronym{BCP}{BCP}{Bayesian Confidence Propagation}
\newacronym{FIFO}{FIFO}{First In First Out}
\newacronym{HBM}{HBM}{High Bandwidth Memory}
\newacronym{PCIe}{PCIe}{Peripheral Component Interconnect Express}
\newacronym{AXI}{AXI}{Advanced eXtensible Interface}
\newacronym{BRAM}{BRAM}{Block Random Access Memory}
\newacronym{DMA}{DMA}{Direct Memory Access}
\newacronym{XRT}{XRT}{Xilinx Runtime library}
\newacronym{SLR}{SLR}{Super Logic Region}
\newacronym{NAISS}{NAISS}{National Academic Infrastructure for Supercomputing in Sweden}
\newacronym{DL}{DL}{Deep Learning}
\newacronym{HCU}{HCU}{hypercolumn unit}
\newacronym{DSP}{DSP}{Digital Signal Processor}
\newacronym{LUT}{LUT}{Look-Up Table}
\newacronym{TRL}{TRL}{Technology Readiness Level}
\newacronym{BLNN}{BLNN}{Brain-Like Neural Network}
\newacronym{PS}{PS}{Processing System}
\newacronym{PL}{PL}{Programmable Logic}
\newacronym{AI}{AI}{Artificial Intelligence}
\newacronym{HPC}{HPC}{High-Performance Computing}

\pgfplotsset{every axis/.append style={
    label style={font=\footnotesize},
    tick label style={font=\footnotesize}
}}

\def\BibTeX{{\rm B\kern-.05em{\sc i\kern-.025em b}\kern-.08em
    T\kern-.1667em\lower.7ex\hbox{E}\kern-.125emX}}
\begin{document}

\title{Embedded FPGA Acceleration of Brain-Like Neural Networks: Online Learning to Scalable Inference
}

\author{\IEEEauthorblockN{1\textsuperscript{st} Muhammad Ihsan Al Hafiz}
\IEEEauthorblockA{\textit{Department of Computer Science} \\
\textit{KTH Royal Institute of Technology}\\
Stockholm, Sweden \\
miahafiz@kth.se}
\and
\IEEEauthorblockN{2\textsuperscript{nd} Naresh Ravichandran}
\IEEEauthorblockA{\textit{Department of Computer Science} \\
\textit{KTH Royal Institute of Technology}\\
Stockholm, Sweden \\
nbrav@kth.se}
\and
\IEEEauthorblockN{3\textsuperscript{rd} Anders Lansner}
\IEEEauthorblockA{\textit{Department of Computer Science} \\
\textit{KTH Royal Institute of Technology} and\\
\textit{Department of Mathematics Stockholm University}\\
Stockholm, Sweden \\
ala@kth.se}
\and
\IEEEauthorblockN{4\textsuperscript{th} Pawel Herman}
\IEEEauthorblockA{\textit{Department of Computer Science} \\
\textit{KTH Royal Institute of Technology}\\
Stockholm, Sweden \\
paherman@kth.se}
\and
\IEEEauthorblockN{5\textsuperscript{th} Artur Podobas}
\IEEEauthorblockA{\textit{Department of Computer Science} \\
\textit{KTH Royal Institute of Technology}\\
Stockholm, Sweden \\
podobas@kth.se}
}

\maketitle

\begin{abstract}
Edge AI applications increasingly require models that can learn and adapt on-device with minimal energy budget. Traditional deep learning models, while powerful, are often overparameterized, energy-hungry, and dependent on cloud connectivity. Brain-Like Neural Networks (BLNNs), such as the Bayesian Confidence Propagation Neural Network (BCPNN), propose a neuromorphic alternative by mimicking cortical architecture and biologically-constrained learning. They offer sparse architectures with local learning rules and unsupervised/semi-supervised learning, making them well-suited for low-power edge intelligence. However, existing BCPNN implementations rely on GPUs or datacenter FPGAs, limiting their applicability to embedded systems. This work presents the first embedded FPGA accelerator for BCPNN on a Zynq UltraScale+ SoC using High-Level Synthesis. We implement both online learning and inference-only kernels with support for variable and mixed precision. Evaluated on MNIST, Pneumonia, and Breast Cancer datasets, our accelerator achieves up to 17.5× latency and 94\% energy savings over ARM baselines, without sacrificing accuracy. This work enables practical neuromorphic computing on edge devices, bridging the gap between brain-like learning and real-world deployment.
\end{abstract}


\section{Introduction}

\glspl{BLNN} are a family of neural networks that follow computational principles of neural information processing in the brain, without sacrificing the ability to practically address computational tasks, e.g., pattern recognition. The general workflow is to use \glspl{BLNN} as a tool to gain insight and model various brain-like functionalities (e.g., learning or cognition) and to later distill them down to be hardware-friendly in order to practically solve problems (e.g., classification problems). One prominent example of a \gls{BLNN} is the \gls{BCPNN} ~\cite{ravichandran_unsupervised_2025}, which has been shown to model brain-like functionality (e.g., odor naming~\cite{chrysanthidis2025elusive}) while at the same time being capable of robust image classification. This model aligns with the goals of neuromorphic computing, which seeks to emulate brain-like computation in a biologically-like and energy-efficient manner, particularly through hardware-software co-design strategies.
 
The number of future edge devices, many of which will contain \gls{AI} capabilities such as \gls{BCPNN}, will grow to be on the orders of billions~\footnote{https://www.statista.com/statistics/1259878/edge-enabled-iot-device-market-worldwide/}. These \gls{BLNN} Edge devices will (in part) take on functionality now in data-centers, and will require to be both compact and be power-efficient to meet societal sustainable development goals~\footnote{https://commission.europa.eu/strategy-and-policy/priorities-2019-2024/europe-fit-digital-age/europes-digital-decade-digital-targets-2030\_en}. Unfortunately, today, \gls{BCPNN} is mainly used in expensive and power-\textit{inefficient} clusters or \gls{HPC} systems, using high-end \glspl{CPU} or \glspl{GPU}, with some prior work adapting \gls{BCPNN} onto expensive data-center \gls{FPGA} accelerator cards~\cite{al_hafiz_reconfigurable_2025}. Thus, in order to meet emerging edge computing demands and facilitate the research and application of \glspl{BLNN} on the Edge, there is a dire need to research and implement new Edge-friendly \gls{BLNN} accelerators.

In this work, we present, to our knowledge, the world's first Edge-capable \gls{BLNN} accelerator for the \gls{BCPNN} model. Our \gls{BLNN} accelerator is built using \gls{HLS} targeting low-power (device sub-Watt) and high-performance (30+ FPS) FPGA acceleration, capable of variable and mixed-precision inference. We claim the following contributions:
\begin{enumerate}
\item The first (to the best of our knowledge) low-power embedded \gls{BCPNN} accelerator for emerging IoT/Edge computing, including design choices, implementation details, and execution strategies,
\item Empirical performance evaluation of our accelerator on three well-known datasets, the impact of variable numerical precision, as well as the positioning of our accelerator against executing on the host ARM A53 processor.
\end{enumerate}
The paper is organized as follows: Section II reviews the background. Section III details our proposed methods. Section IV covers implementation. Section V presents experimental results, and Section VI discusses related work. Section VII provides the conclusion.

\section{Background}

\subsection{Brain-Like Neural Networks}

\glspl{BLNN} are computational models more strictly constrained by the architecture and function of the human brain than mainstream biologically inspired neural networks. For instance, they rely on sparse connectivity and use correlation-based Hebbian types of learning instead of any form of error backpropagation. A prominent example is the \gls{BCPNN} ~\cite{ravichandran_unsupervised_2025}. This model specifically mimics the modularization of the primate neocortex in terms of hypercolumns and functional minicolumns and integrates Bayesian-Hebbian learning and inference with structural plasticity for activity-dependent rewiring of sparse connectivity~\cite{ravichandran_unsupervised_2025, mountcastle_columnar_1997}. This biologically inspired approach inherently supports cognitive abilities such as unsupervised and semi-supervised learning, associative memory functionality, and is capable of reward-based learning~\cite{hawkins_theory_2017,ravichandran_brain-like_2021}. This brain-like architecture featuring local in-memory learning and computation provides a hardware-friendly and robust function with little tendency for overfitting of training data.

\begin{figure}[!ht]
    \centering
    \includegraphics[width=0.8\linewidth]{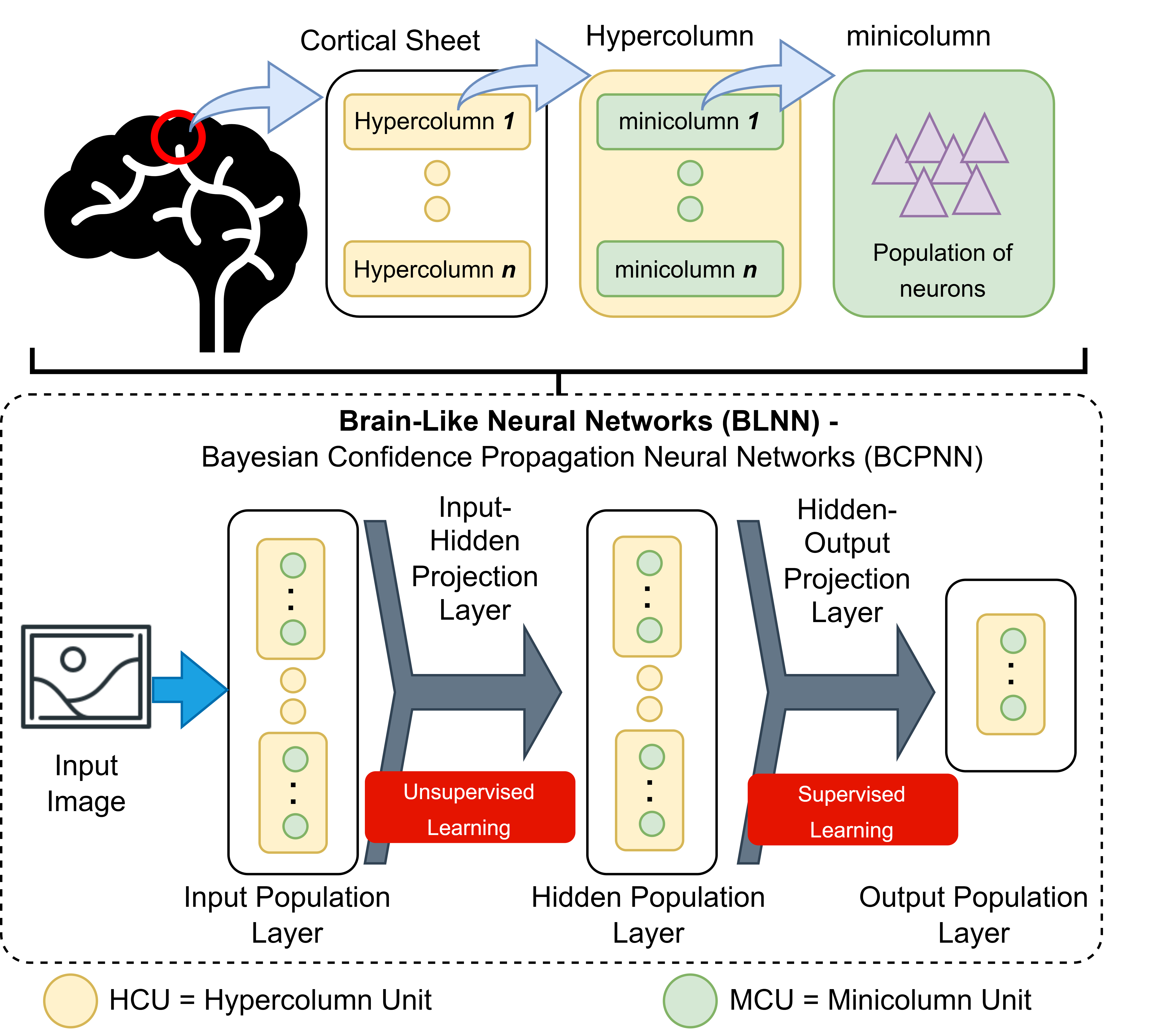}
    \caption{Architecture of a BLNN based on the BCPNN model, featuring HCU and MCU. Learning occurs in two phases: unsupervised in the input-hidden layer and supervised in the hidden-output layer.}
    \label{fig:blnn}
\end{figure}

The cortical architecture adopted by \gls{BCPNN} features Populations (arrays of neural units representing, e.g., a cortical area) and Projections (representing bundles of connections between such areas) as the main building blocks. Each population is further modularized in terms of soft-winner-take-all Hypercolumn Units (HCUs), each comprising several Minicolumn Units (MCUs). Each  MCU represents a population of neurons in the biological neocortex. Altogether, these components collectively form a network-of-network architecture that resembles the modular organization of the primate neocortex~\cite{mountcastle_columnar_1997,hawkins_theory_2017}. Figure~\ref{fig:blnn} illustrates the fundamental structure and modular design of \glspl{BLNN} explored in this paper.

 Structurally, the \gls{BLNN} model developed in this project consists of three populations ("layers"): the \textit{input layer}, the \textit{hidden layer}, and the \textit{output layer}. The populations in these layers have HCUs organized into MCUs normalized by means of soft-WTA dynamics, and the input-to-hidden layer projection features structural plasticity, which rewires this connectivity to form efficient, sparse activity internal representations in the hidden layer ~\cite{ravichandran_brain-like_2021, ravichandran_unsupervised_2025}. Sparseness in network connectivity is controlled by the amount of active ($n_\text{act}$) synapses and the amount of silent synapses ($n_\text{sil}$) supports replacement of not so useful synapses with new synapses which are yet not present in the connectivity between the input and the projection layer. This data-dependent rewiring process is an important property of structural \gls{BCPNN} plasticity, allowing end users to control the communication and computation demand and pattern processing capabilities of the network model~\cite{ravichandran_brain-like_2021}. 

\gls{BCPNN} leverages biologically plausible local feedforward learning mechanisms without relying on error backpropagation~\cite{ravichandran_unsupervised_2025}. The learning process consists of two distinct phases: an unsupervised phase in the input-to-hidden projection layer, followed by a supervised phase in the hidden-to-output projection layer. During the unsupervised phase, input patterns propagate from the input layer to the hidden layer, where hidden hypercolumns form receptive fields to capture input features~\cite{ravichandran_unsupervised_2025}. Subsequently, the hidden-to-output projection layer employs supervised learning, associating the formed hidden-layer representations with labeled output data, ultimately enabling classification or other inference tasks.

Central to the learning dynamics of \gls{BCPNN} models are probability-based synaptic variables, known as \textit{synaptic traces} (usually z-,  p-traces, sometimes e-traces for delayed reward learning). These traces encode the statistical probabilities in the form of exponential moving averages of neural activation events, with different time constants~\cite{ravichandran_unsupervised_2025}. In this paper, the spike activity is represented as rate-based and not as an actual spike. The p-traces track three probabilities: the probability of pre-synaptic MCU activation ($p_{i}$), the probability of post-synaptic MCU activation ($p_{j}$), and the joint probability of simultaneous activation of both pre- and post-synaptic MCUs ($p_{ij}$). These traces are updated continuously and incrementally with a predefined learning rate $\alpha$~\cite{ravichandran_unsupervised_2025}.

The Bayesian-Hebbian learning rule of \gls{BCPNN} computes biases and connection weights directly from these probability traces~\cite{ravichandran_unsupervised_2025,hawkins_theory_2017}. Specifically, the bias of a hidden MCU ($b_{j}$) encodes the self-information or prior:

\begin{equation}
b_{j} = \log p_{j}
\end{equation}

Similarly, the synaptic weight between an input MCU ($x_{i}$) and a hidden MCU ($y_{j}$) encodes their statistical dependency through the ratio of their joint probability and the product of their individual activation probabilities, i.e., point-wise mutual information:

\begin{equation}
w_{ij} = \log\left(\frac{p_{ij}}{p_{i}\cdot p_{j}}\right)
\end{equation}

This probabilistic interpretation provides a robust computational framework for various learning tasks, with implications for neuromorphic computing and embedded neural network implementations.
Currently, \gls{BCPNN} is primarily implemented on CPUs, GPUs \cite{ravichandran_unsupervised_2025}, or high-performance FPGAs \cite{wang_fpga-based_2024,wang_scalable_2025,al_hafiz_reconfigurable_2025}, with limited exploration on embedded platforms for edge-device acceleration.

\subsection{Embedded-FPGA Design Constraints}

Embedded FPGAs, also known as a System-On-Chip (SoC), are a type of FPGA that is physically embedded in similar chip packaging with other components, for instance, \gls{CPU} or Processing System (PS), digital interfaces (such as SPI, I2C, UART), analog interfaces, memory controllers, and other peripherals \cite{collini_reconfigurable_2022}. This integration brings benefits in power, performance, practicality, and cost, due to reduced inter-component communication, compact form factor, efficient resource sharing, and the ability to offload tasks to custom logic for improved energy and execution efficiency. An example of a commonly used embedded FPGA is the Zynq UltraScale+ from AMD/Xilinx. The Zynq UltraScale+ integrates an ARM processor along with FPGA fabric and other peripherals, offering a comprehensive platform for embedded computing and real-time processing. The embedded FPGA is commonly used in broad engineering applications such as robotics \cite{basha_implementation_2022}, machine learning \cite{tsai_-chip_2023}, healthcare \cite{liu_fpga-based_2020}, and so on. However, embedded \glspl{FPGA} bring the challenge of less memory (both external and on-chip), fewer hardware resources, and a tighter power budget \cite{garcia_optimized_2019, collini_reconfigurable_2022}.

\begin{table}[!htbp]
    \centering
    \caption{Comparison between embedded FPGA (ZCU104) and datacenter FPGA (Alveo U55C).}
    \label{tab:platform_comparison}
    \renewcommand{\arraystretch}{1.05}
    \begin{tabular}{c|c|c}
        \hline
        \textbf{Parameter} & \textbf{ZCU104 (XCZU7EV)} & \textbf{Alveo U55C (XCU55C)} \\
        \hline \hline
        Type & SoC (Embedded FPGA) & Accelerator Card \\
        Usage Target & Edge / Real-Time & Cloud / HPC \\
        LUTs & 230400 & 1303680 \\
        Flip-Flops (FFs) & 460800 & 2607360 \\
        DSP Slices & 1728 & 9024 \\
        BRAM & 11 Mb & 70.9 Mb \\
        Off-Chip Memory & 2 GB DDR4 (PS-side) & 16 GB HBM2 \\
        Max Power Budget & $<30 W$& $150W$ \\
        \hline
    \end{tabular}
\end{table}

One of the commonly used embedded FPGAs is ZCU104 \cite{reddy_fpgazcu104_2024, tsai_fpga-based_2025}. Table \ref{tab:platform_comparison} compares the embedded FPGA ZCU104 against a data-center FPGA accelerator card: the Alveo U55C. The table highlights the significantly more constrained compute fabric, memory, and power envelope in embedded platforms. While datacenter FPGAs offer abundant resources for high-throughput parallel processing, embedded FPGAs must deliver efficient performance within tight area and energy limits. These constraints necessitate carefully optimized architectural design choices, particularly in resource allocation, parallelization, and numerical precision. 
In the following section, we present our system-level architecture and methods that are explicitly tailored to address these embedded FPGA constraints while maintaining functionality and efficiency.

\section{Methods}

\subsection{System-Level Overview}
\label{sec:system-level-overview}

The system-level implementation of our \gls{BLNN} accelerator targets a heterogeneous embedded FPGA platform, which integrates a Processing System (PS) or CPU, and Programmable Logic (PL) or FPGA fabric. Our design leverages tight PS-PL integration, optimizing data movement and parallel computation across hardware and software. We implemented a \gls{BCPNN} model comprising three layers: input, hidden, and output, interconnected via two projections (input-hidden and hidden-output). Two kernel variants were developed: a \textit{full online-learning kernel} supporting dynamic synaptic plasticity, and an \textit{inference-only kernel} optimized for low-power edge deployment. The key functional difference is the presence of synaptic trace and weight-bias update operations, which are exclusive to the full online-learning kernel.

\begin{figure}[!ht]
    \centering
    \includegraphics[width=0.9\linewidth]{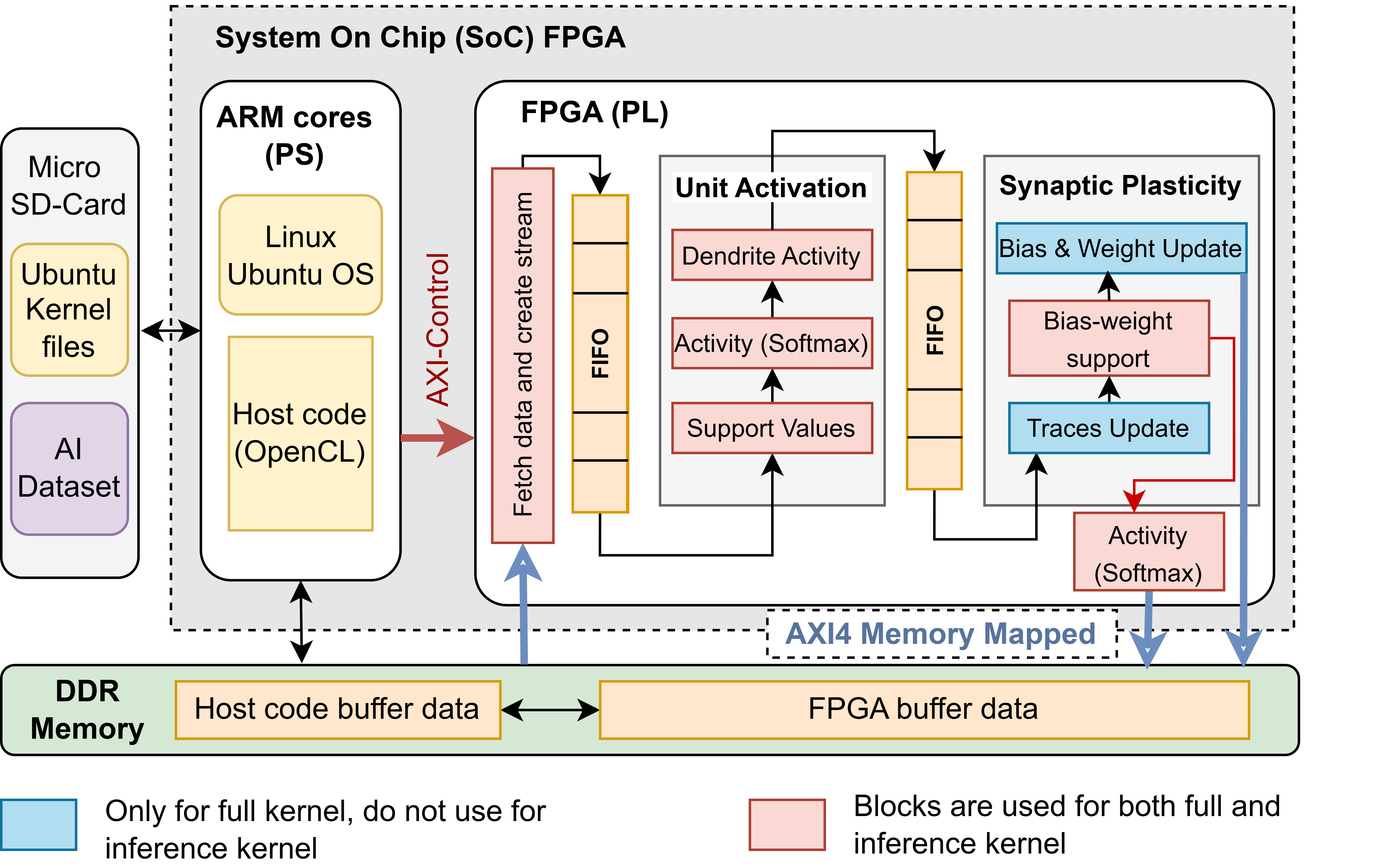}
    \caption{BCPNN system architecture on embedded FPGA.}
    \label{fig:block-diagram}
\end{figure}

Figure~\ref{fig:block-diagram} provides the \gls{BLNN} system architecture illustration on an embedded FPGA. The kernels derive from the latest \gls{BCPNN} theory \cite{ravichandran_unsupervised_2025} and high-performance FPGA implementations \cite{al_hafiz_reconfigurable_2025}, adapted here for embedded FPGA constraints. ARM cores as PS run Ubuntu OS to handle the orchestrations. The host code, deployed on the PS side, fetches the dataset from the Micro SD card to the DDR memory. At the same time, the buffer memory for the FPGA is prepared in DDR. The host code initiates the run of the kernel in the FPGA fabric or PL via AXI control. After the operation is started in the kernel, it fetches data from DDR via dedicated 256-bit AXI4 Memory-Mapped bursts (8 float precision or 16 half precision values per cycle) and converts it into a stream. Some of the stream variables are packed with an array (8 or 16, depending on the precision), and some are in single data, depending on the parallel potential operation in the sub-kernel. Sub-kernels are interconnected via AXI-stream FIFO channels, enabling dataflow and parallel execution with a data-driven process. Within the sub-kernel, the loop-unroll pragma sets the internal parallel factor.

\subsection{Full Online-Learning Kernel}
\label{sec:full-online-learning-kernel}

The full kernel retains the stream-based pipeline of \cite{al_hafiz_reconfigurable_2025} but scales it to an embedded device by capping the unroll factor at 4 due to limited hardware resources. With that width, the costly data-merging FIFOs of the original design are unnecessary, which saves BRAM utilization. Vitis-\gls{HLS} \textit{unsafe-math} is enabled for aggressive floating-point optimizations to reduce resource usage, increase performance, and shorten latency. 

To support parallel dataflow during the fetch data process, most kernel input-output parameters are mapped to separate AXI interfaces. However, less time-critical variables, such as inputs, labels, random values, and outputs, are bundled into a single AXI interface, as their sequential access does not hinder performance. This consolidation, validated via \gls{RTL} co-simulation, effectively reduces \gls{BRAM} utilization. For the input-to-hidden weight matrix, which demands a deep FIFO, we adopt a specialized solution. Since the weight is accessed concurrently by multiple subkernels (bias-weight support update and weight update), we assign multiple AXI bundles to enable burst fetching and generate multiple parallel streams, thereby avoiding deadlocks and minimizing FIFO depth requirements.


\subsection{Inference Kernel with Precision Variants}
\label{sec:inference-precision-variants}

The inference-only kernel omits synaptic trace and bias-weight update operations, significantly reducing FPGA resource requirements (DSP, BRAM, LUT, registers). Consequently, it supports greater computational parallelism enabled by full-width 256-bit burst access. To evaluate the trade-offs in accuracy, speed, resource use, and energy efficiency, we implemented three distinct precision variants:

\paragraph{Float Precision (FP32, baseline)}
This variant employs standard 32-bit IEEE-754 floating-point arithmetic, fetching eight data points per burst cycle. While similar in accuracy to the full kernel's inference mode, the inference-only variant achieves superior speed due to higher burst parallelism (factor of 8 versus 4 in the full kernel).

\paragraph{Half Precision (FP16)}
The FP16 variant doubles the parallel data fetch to sixteen values per cycle. So, it can increase the parallelism factor to 16. With a numeric range of approximately $\pm65,504$ and a minimum positive increment of $1/1024$, FP16 offers sufficient numeric precision for \gls{BLNN} inference, substantially reducing FPGA resource usage compared to FP32.

\paragraph{Mixed Precision (FP16 \& FXP16)}
To further improve resource efficiency, we explored mixed precision using fixed-point (16-bit, Q3.12 format: 4 integer bits, 12 fractional bits) for data storage and FP16 for computationally intensive operations requiring larger numeric ranges, such as accumulations. The Q3.12 provides values in the range $\pm8$ with a minimum increment of $2^{-12}$, and enough to represent the trained parameters. The use of FP16 for accumulation prevents overflow and maintains accuracy. This configuration retains the burst parallelism factor of 16.

\subsection{Trained parameter flow}

\begin{figure}[!ht]
    \centering
    \includegraphics[width=0.8\linewidth]{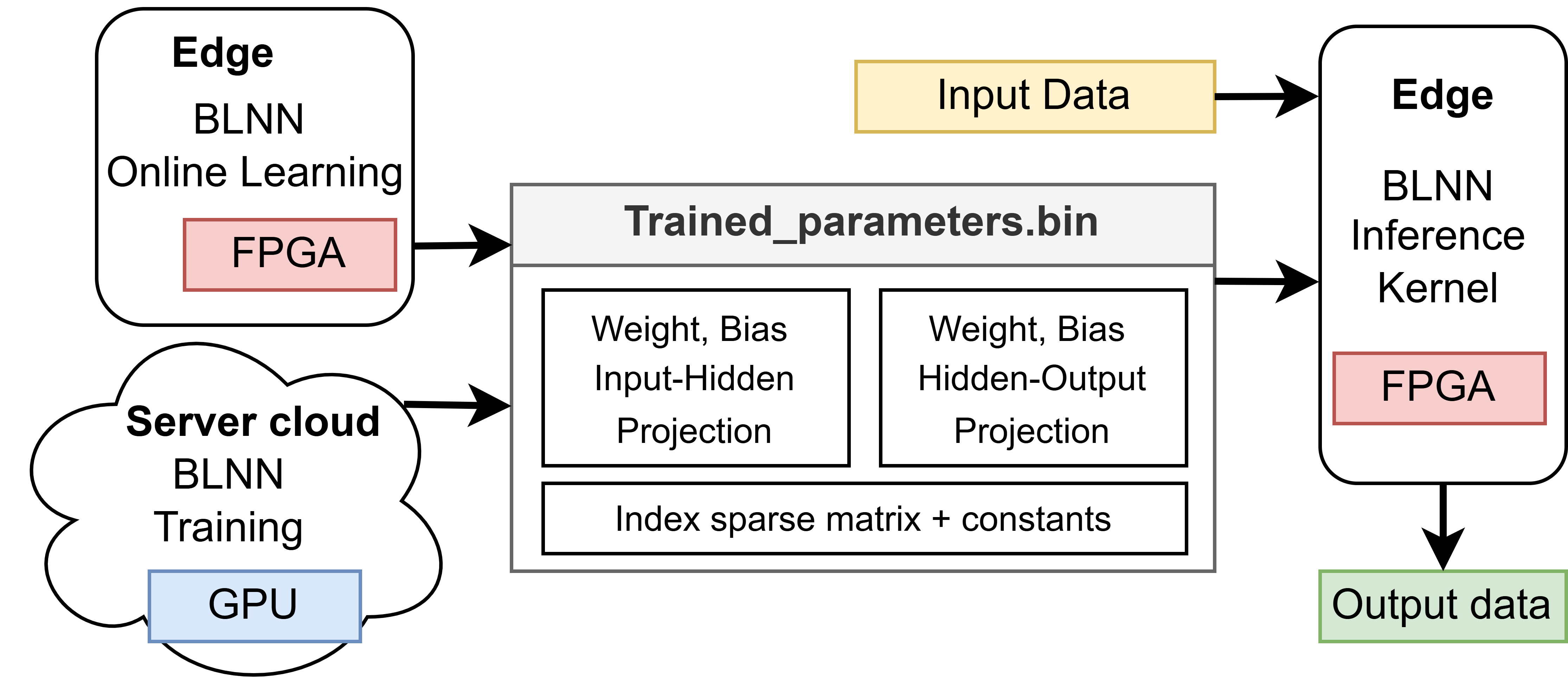}
    \caption{Inference workflow for BLNN.}
    \label{fig:training-flow}
\end{figure}

Figure~\ref{fig:training-flow} depicts the workflow from training to inference deployment. Training produces a binary file containing model parameters: weights, biases, sparse indices, and constants. \gls{BLNN} training is platform-agnostic and can occur on high-performance platforms (cloud GPUs) or on FPGA hardware. For inference deployment, this parameter file is transferred once into the FPGA buffer memory, significantly reducing subsequent runtime host interactions and energy overhead.

\section{Implementation}

\subsection{Hardware \& Toolchain}

All experiments were run on an AMD Xilinx ZCU104 SoC-FPGA board (XCZU7EV) under Ubuntu 22.04.
The host program, written in C++, uses OpenCL to launch kernels in the PL or FPGA fabric. The FPGA configuration and management leveraged the Xilinx-provided \texttt{xlnx-config} command-line tool, enabling hardware platform initialization, bitstream management, and device configuration.

Our \gls{BCPNN} kernel is written in C++ and synthesized with Vitis 2023.2 \gls{HLS}. Our implementation methodology employed the Vitis workflow, where we first generated an .xo file through \gls{HLS}, and then used it in the system project to perform synthesis and implementation, ultimately generating the FPGA bitstream. During compilation, we employed \texttt{config\_compile -unsafe\_math\_optimizations} to enable the unsafe math optimization and \texttt{config\_interface -m\_axi\_max\_widen\_bitwidth 256} to allow AXI interfaces to utilize burst transfers with a maximum data width of 256 bits. Moreover, the clock frequency for the FPGA kernel was tuned to avoid the worst critical path and timing violation: 120 MHz for the float precision kernel and 130 MHz for half precision and mixed-precision variants. RTL co-simulation guided FIFO-depth choices, guaranteeing deadlock-free AXI-Stream traffic while minimizing BRAM.


\subsection{Datasets \& Model Sizes}
\label{sec:dataset}

We evaluated our \gls{BCPNN} FPGA implementation using three public datasets: MNIST\cite{lecun_gradient-based_1998}, Pneumonia\cite{yang_medmnist_2023}, and Breast Cancer\cite{yang_medmnist_2023}, to demonstrate the model's versatility across a range of data types and classification challenges. Table~\ref{tab:model} summarizes the datasets, configurations, and training parameters.


\begin{table}[!ht]
\centering
\caption{Model Configurations for Different Datasets and Kernels}
\label{tab:model}
\begin{tabular}{l|c|c|c}
\hline
\textbf{Parameter} & \textbf{MNIST} & \textbf{Pneumonia} & \textbf{Breast Cancer} \\
\hline
Kernel(s)           & Full, Infer.     & Inference only     & Inference only \\
In/Out dims         & ($28\times28$)/10  & ($64\times64$)/2     & ($128\times128$)/2 \\
HCU/MCU             & 32/128           & 10–30 / 200–400    & 10/1000 \\
$n_\text{act}/n_\text{sil}$ & 64 / 64     & 80–320 / 24–80     & 676 / 156 \\
Epoch / $\tau_p$    & 5 / 3            & 5 / 0.3            & 15 / 0.2 \\
\hline
\end{tabular}
\end{table}

The \textit{MNIST} dataset, comprising handwritten digits (0-9) with $28\times28$ images, was evaluated using both the full online-learning kernel and the inference kernel. The \textit{Pneumonia} dataset, containing chest X-ray images resized to $64\times64$, tested the inference kernel with varying configurations of HCU, MCU, and connectivity sparsity ($n_\text{act}, n_\text{sil}$). The \textit{Breast Cancer} dataset evaluated the inference kernel with larger $128\times128$ images, demonstrating the embedded FPGA’s capacity to handle larger input data while maintaining accuracy.

In Table~\ref{tab:model}, the row \textit{HCU/MCU} defines hidden-layer sizes. The 
$n_\text{act}$ and $n_\text{sil}$ parameters specify connectivity sparsity between input and hidden layers in terms of of number of active and silent synaptic connections per hidden HCU. \textit{Epoch} indicates the number of training iterations through the full dataset, while \textit{$\tau_p$} controls exponential decay and integration of neural activity traces, influencing network memory and learning dynamics. Higher $\tau_p$ increases temporal memory, while smaller values provide faster adaptation to recent input patterns, like in short-term memory. Furthermore, the model parameters were selected through empirical iteration to balance classification accuracy and hardware efficiency. These configurations also represent a broad design space to evaluate scalability across varying resource budgets. Notably, the model remains reconfigurable, allowing parameter tuning to suit different application requirements.

\subsection{Measurement Methodology}
\label{sec:method_measure}

We evaluated the \gls{BCPNN} FPGA implementation across three metrics: latency, power consumption, and accuracy.

\subsubsection{Latency Measurement}
Execution latency was measured using the \texttt{gettimeofday()} POSIX system call, which provides microsecond-resolution timestamps. Timestamps captured immediately before and after kernel execution on the host, including the memory transfers, encompass complete kernel execution on the FPGA.

\subsubsection{Power Measurement}
Power consumption measurements employed the onboard INA226 power-monitoring IC on the ZCU104 board. We acquired two power traces: the board and execution power. Board power is the overall power that is required from the whole board. Execution power is the difference between board power during idle and execution. Hence, the true power consumption of our accelerator lies between the board power (that also includes unused components of our development board) and the Execution power (which only counts the dynamic power). We sampled instantaneous power consumption from the main input line at 10 ms intervals to capture potential fluctuating power, recording more than 1000 samples each for idle and active kernel execution states. This method accurately accounts for power overheads from data transfers and host-PL interactions.

\subsubsection{Accuracy Measurement}
Accuracy was evaluated exclusively using test data (unseen during training). Predicted classifications from the FPGA implementation were directly compared to ground-truth labels, generating reliable accuracy metrics indicative of real-world inference performance.

\section{Results}

\subsection{Baseline Performance (Float Precision)}

The baseline evaluation compares latency, accuracy, power, and FPGA resource utilization of our \gls{BCPNN} implementations (full and inference-only kernels) against an ARM A53 core baseline in float precision with similar model and execution parameters.

\begin{table}[ht]
\centering
\caption{Comparison of Latency and Accuracy for ARM and FPGA}
\label{tab:latency_speedup_accuracy}
\renewcommand{\arraystretch}{1.2}
\setlength{\tabcolsep}{4pt}
\begin{tabular}{>{\centering\arraybackslash}p{1.7cm}|
                >{\centering\arraybackslash}p{1.2cm}|
                >{\centering\arraybackslash}p{1.6cm}|
                >{\centering\arraybackslash}p{1.6cm}|
                >{\centering\arraybackslash}p{1.3cm}}
\hline 
\textbf{Dataset} & \textbf{Platform} & \textbf{Latency (ms)} & \textbf{Speed-up} & \textbf{Accuracy} \\
\textbf{(kernel)} & & \textbf{Train / Infer} & \textbf{over ARM} & \textbf{Test (\%)} \\
\hline \hline
\multirow{2}{*}{\shortstack{MNIST\\(Full Kernel)}} 
    & ARM  & 145.5 / 36.7  &       & 94.6 \\
    & FPGA & 55.4 / 17.8  & 2.63x / 2.06x  & 94.3 \\
\hline
\multirow{2}{*}{\shortstack{MNIST\\(Infer. Kernel)}} 
    & ARM  & - / 37.8  &       & 94.6 \\
    & FPGA & - / 3.4  & - / 11.12×  & 94.6 \\
\hline
\multirow{2}{*}{\shortstack{Pneumonia\\(Infer. Kernel)}} 
    & ARM  & - / 309.1  &       & 86.2 \\
    & FPGA & - / 18.8  & - / 16.45×  & 86.2 \\
\hline
\multirow{2}{*}{\shortstack{Breast\\(Infer. Kernel)}} 
    & ARM  & - / 531.9  &       & 84.0 \\
    & FPGA & - / 30.3  & - / 17.56×  & 84.0 \\
\hline
\end{tabular}
\end{table}
Table~\ref{tab:latency_speedup_accuracy} summarizes latency, speed-up, and accuracy results. The general trend is that the larger the network that is inferred, the more benefit our FPGA accelerator shows over the ARM baseline, as shown empirically with MNIST (smallest) achieving 2.6x compared to Breast (largest) achieving 17.56x. The full kernel shows lower speed-ups (2.06×-2.63×), primarily due to it requiring more subkernels to run and limited parallelization (factor of 4), compared to a factor of 8 and fewer subkernels to run in the inference-only kernel. Accuracy remains consistent between ARM and FPGA implementations, with slight variance in the full kernel likely attributed to minor differences in random initialization and "unsafe-math" compile properties.

\begin{table}[ht]
\centering
\caption{Comparison of Power and Energy for ARM and FPGA}
\label{tab:power_energy}
\renewcommand{\arraystretch}{1.2}
\setlength{\tabcolsep}{4pt}
\begin{tabular}{>{\centering\arraybackslash}p{1.6cm}|
                >{\centering\arraybackslash}p{1cm}|
                >{\centering\arraybackslash}p{1.7cm}|
                >{\centering\arraybackslash}p{1.7cm}|
                >{\centering\arraybackslash}p{1.4cm}}
\hline 
\textbf{Dataset} & \textbf{Platform} & \textbf{Power (W)} & \textbf{Energy (mJ)} & \textbf{Energy} \\
\textbf{(kernel)} & & \textbf{Board / Exec.} & \textbf{Board / Exec.} & \textbf{Saving (\%)} \\
\hline \hline
\multirow{2}{*}{\shortstack{MNIST\\(Full Kernel)}} 
    & ARM  & 11.134 / 0.222  & 1620.2 / 32.3   &  \\
    & FPGA & 11.447 / 0.558  & 633.7 / 30.9  & 60.9 / 4.4 \\
\hline
\multirow{2}{*}{\shortstack{MNIST\\(Infer. Kernel)}} 
    & ARM  & 10.388 / 0.241  & 392.8 / 9.1     &  \\
    & FPGA & 10.476 / 0.299  & 35.1 / 1.0  & 91.1 / 89.0 \\
\hline
\multirow{2}{*}{\shortstack{Pneumonia\\(Infer. Kernel)}} 
    & ARM  & 10.385 / 0.238  & 3209.6 / 73.6     &  \\
    & FPGA & 10.728 / 0.481  & 201.2 / 9.0  & 93.7 / 87.7 \\
\hline
\multirow{2}{*}{\shortstack{Breast\\(Infer. Kernel)}} 
    & ARM  & 10.411 / 0.264  & 5537.7 / 140.4     &  \\
    & FPGA & 10.803 / 0.564  & 327.7 / 17.1  & 94.1 / 87.8 \\
\hline
\end{tabular}
\end{table}

Table~\ref{tab:power_energy} presents board and execution-specific power and energy metrics. FPGA implementations exhibit slightly higher power due to FPGA fabric usage. However, the substantial latency reductions translate directly into energy savings. The inference-only kernels deliver notable energy savings (up to 94.1\%) compared to ARM. The full kernel also reduces energy but less dramatically (up to 60.9\%), reflecting its higher computational complexity and lower parallelism.

\begin{figure}[htbp]
\centering
\begin{adjustbox}{max width=\textwidth}
\begin{tikzpicture}
\begin{axis}[
    ybar,
    bar width=5pt,
    width=8cm, height=5cm,
    enlarge x limits={abs=0.6cm},
    ylabel={\footnotesize Resource Usage (\%)},
    ylabel style={yshift=-5pt}, 
    symbolic x coords={MNIST full (training), MNIST inference, Pneumonia inference, Breast inference},
    xtick=data,
    xticklabels={
        {\shortstack{MNIST\\full (training)}},
        {\shortstack{MNIST\\inference}},
        {\shortstack{Pneumonia\\inference}},
        {\shortstack{Breast\\inference}}
    },
    xticklabel style={align=center, yshift=0pt,rotate=0, font=\footnotesize},
    ymin=0,
    ymax=100,
    ymajorgrids=true,
    legend style={at={(0.5,-0.25)}, anchor=north, legend columns=4},
    nodes near coords,
    nodes near coords style={
        rotate=90,
        anchor=west,
        font=\scriptsize
    },
    legend style={
        at={(0.98,0.98)},
        anchor=north east,
        draw=black,
        fill=white,
        font=\footnotesize
    }
]

\addplot+[fill=blue]    coordinates {(MNIST full (training),49.40) (MNIST inference,18.00) (Pneumonia inference,18.53) (Breast inference,18.45)};
\addplot+[fill=red]     coordinates {(MNIST full (training),39.88) (MNIST inference,11.19) (Pneumonia inference,11.56) (Breast inference,11.2)};
\addplot+[fill=green]   coordinates {(MNIST full (training),31.37) (MNIST inference,13.43) (Pneumonia inference,13.48) (Breast inference,13.48)};
\addplot+[fill=orange]  coordinates {(MNIST full (training),76.28) (MNIST inference,25.16) (Pneumonia inference,39.58) (Breast inference,51.44)};

\legend{LUT, FF, DSP, BRAM}
\end{axis}
\end{tikzpicture}
\end{adjustbox}
\caption{FPGA Resource Usage for Baseline BLNN Kernels float precision}
\label{fig:fpga_resources}
\end{figure}
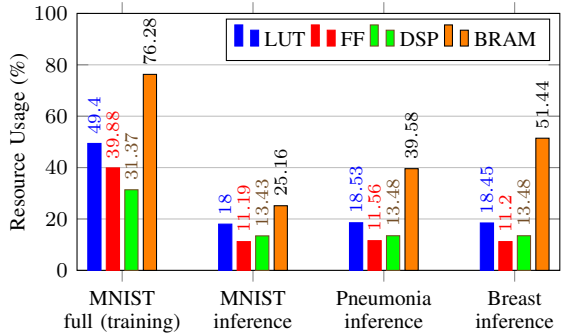

Figure~\ref{fig:fpga_resources} illustrates FPGA resource utilization. The full kernel consumes significantly higher BRAM (76.28\%) because it has more AXI interfaces to transfer more variables for the learning mechanism, and local variables in subkernel synaptic plasticity. Although DSP utilization (31.37\%) could theoretically support increased parallelism, synthesis trials indicated that higher parallelization would substantially increase LUT and FF usage, causing placement and routing challenges. Thus, the full kernel parallelism factor was limited to 4. In contrast, inference-only kernels consistently show lower utilization across resources (LUT, FF, DSP). These results suggest substantial headroom remains for scaling inference kernels to even larger models.

\subsection{Precision Sensitivity Analysis}

We evaluated three precision variants for our inference-only kernels: float (FP32, baseline), half (FP16), and mixed precision (FXP16/FP16).

\begin{figure}[htbp]
\centering
\begin{tikzpicture}
\begin{axis}[
    width=1\columnwidth,
    height=5cm,
    grid=major,
    xlabel={Latency (ms)},
    ylabel={Energy (mJ)},
    enlargelimits=true,
    scatter/classes={
        MNIST_float={mark=*,draw=black,fill=blue},
        MNIST_half={mark=*,draw=black,fill=red},
        MNIST_mixed={mark=*,draw=black,fill=green},
        PNEU_float={mark=square*,draw=black,fill=blue},
        PNEU_half={mark=square*,draw=black,fill=red},
        PNEU_mixed={mark=square*,draw=black,fill=green},
        BREAST_float={mark=triangle*,draw=black,fill=blue},
        BREAST_half={mark=triangle*,draw=black,fill=red},
        BREAST_mixed={mark=triangle*,draw=black,fill=green}
    },
    legend style={at={(0.5,-0.25)}, anchor=north, legend columns=3, font=\scriptsize},
]

\addplot[scatter,only marks,mark size=3.5pt,scatter src=explicit symbolic]
coordinates {
    (3.354, 1.003) [MNIST_float]
    (2.29, 0.573) [MNIST_half]
    (2.293, 0.601) [MNIST_mixed]
    
    (18.759, 9.023) [PNEU_float]
    (10.602, 4.697) [PNEU_half]
    (10.85, 4.774) [PNEU_mixed]
    
    (30.333, 17.108) [BREAST_float]
    (16.236, 9.271) [BREAST_half]
    (16.215, 9.145) [BREAST_mixed]
};

\legend{
MNIST-Float (94.6\%), MNIST-Half (94.6\%), MNIST-Mixed (94.7\%),
Pneu-Float (86.2\%), Pneu-Half (86.4\%), Pneu-Mixed (84.1\%),
Breast-Float (84.0\%), Breast-Half (84.0\%), Breast-Mixed (78.8\%)
}

\end{axis}
\end{tikzpicture}
\caption{Latency vs Power across Datasets and Precision Types. Accuracy is indicated in the legend.}
\label{fig:2d_latency_power}
\end{figure}
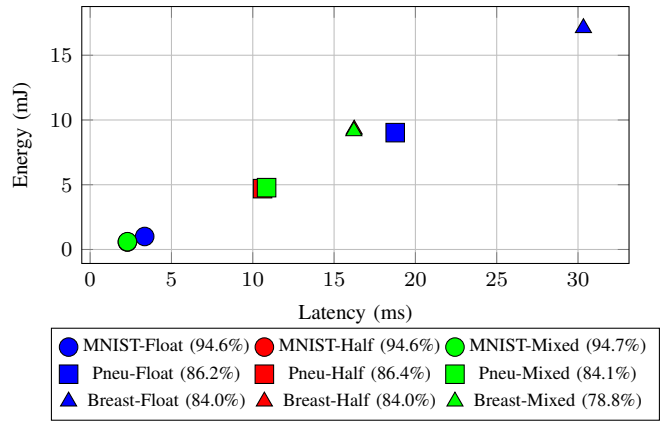

Figure~\ref{fig:2d_latency_power} compares latency and FPGA execution energy across precisions. Both half and mixed precision significantly reduce latency and energy across datasets. This improvement results from the higher parallelization factor (16 data elements per memory fetch for FP16/FXP16 vs. 8 for FP32), improving throughput with equivalent bandwidth while reducing energy and latency. Moreover, Accuracy is stable when precision reduces from float to half for all datasets (e.g., MNIST remains at 94.6\%, Pneumonia slightly improves to 86.4\%, Breast remains at 84\%). However, mixed precision introduces minor accuracy loss for more complex datasets (Pneumonia drops to 84.1\%, Breast notably to 78.8\%). MNIST accuracy remains unaffected, likely due to its lower input complexity and lower sensitivity to numeric precision.

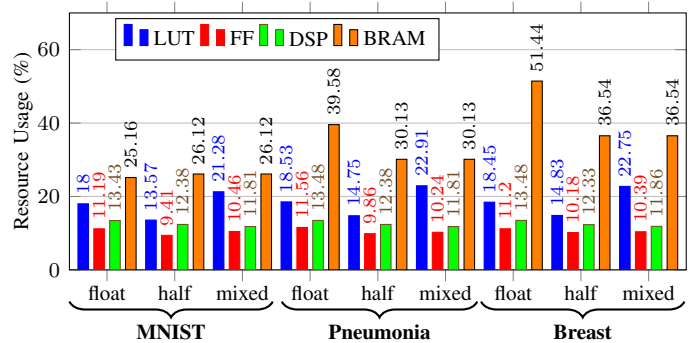
\begin{figure}[htbp]
\centering
\begin{adjustbox}{max width=\textwidth}
\begin{tikzpicture}
\begin{axis}[
    ybar,
    bar width=4pt,
    x=0.9cm, 
    width=20cm, height=5cm,
    enlarge x limits={abs=0.5cm},
    ylabel={\footnotesize Resource Usage (\%)},
    ylabel style={yshift=-5pt}, 
    symbolic x coords={mnist float, mnist half, mnist mixed, 
    pneu float, pneu half, pneu mixed,
    breast float, breast half, breast mixed},
    xtick=data,
    xticklabels={float, half, mixed, float, half, mixed, float, half, mixed},
    xticklabel style={align=center, yshift=2pt,rotate=0, font=\footnotesize},
    ymin=0,
    ymax=70,
    ymajorgrids=true,
    legend style={at={(0.5,-0.25)}, anchor=north, legend columns=4},
    nodes near coords,
    nodes near coords style={
        rotate=90,
        anchor=west,
        font=\scriptsize
    },
    legend style={
        at={(0.6,0.98)},
        anchor=north east,
        draw=black,
        fill=white,
        font=\footnotesize
    }
]

\addplot+[fill=blue]    coordinates {(mnist float,18.00) (mnist half,13.57) (mnist mixed,21.28) (pneu float,18.53) (pneu half,14.75) (pneu mixed,22.91) (breast float,18.45) (breast half,14.83) (breast mixed,22.75) };
\addplot+[fill=red]     coordinates {(mnist float,11.19) (mnist half,9.41) (mnist mixed,10.46) (pneu float,11.56) (pneu half,9.86) (pneu mixed,10.24) (breast float,11.20) (breast half,10.18) (breast mixed,10.39) };
\addplot+[fill=green]   coordinates {(mnist float,13.43) (mnist half,12.38) (mnist mixed,11.81) (pneu float,13.48) (pneu half,12.38) (pneu mixed,11.81) (breast float,13.48) (breast half,12.33) (breast mixed,11.86) };
\addplot+[fill=orange]  coordinates {(mnist float,25.16) (mnist half,26.12) (mnist mixed,26.12) (pneu float,39.58) (pneu half,30.13) (pneu mixed,30.13) (breast float,51.44) (breast half,36.54) (breast mixed,36.54) };
\legend{LUT, FF, DSP, BRAM}
\end{axis}

\draw[decorate,decoration={brace, mirror, amplitude=6pt}, thick, yshift=-12pt]
    ([xshift=0pt]0,0) -- ([xshift=20pt]2,0) node[midway, below=5pt, font=\footnotesize]{\textbf{MNIST}};

\draw[decorate,decoration={brace, mirror, amplitude=6pt}, thick, yshift=-12pt]
    ([xshift=-5pt]3,0) -- ([xshift=12pt]5,0) node[midway, below=5pt, font=\footnotesize]{\textbf{Pneumonia}};

\draw[decorate,decoration={brace, mirror, amplitude=6pt}, thick, yshift=-12pt]
    ([xshift=-15pt]6,0) -- ([xshift=5pt]8,0) node[midway, below=5pt, font=\footnotesize]{\textbf{Breast}};

\end{tikzpicture}
\end{adjustbox}
\caption{FPGA Resource Usage for precision variants}
\label{fig:precision_variant_util}
\end{figure}

Figure~\ref{fig:precision_variant_util} shows FPGA resource utilization across precision formats. BRAM usage decreases significantly for large models (Pneumonia, Breast) when precision is lowered, but slightly increases for MNIST due to additional local variables associated with higher parallelization. LUT and FF usage drop from float to half precision, but rise slightly in mixed precision due to format conversion overheads. DSP usage is modestly reduced in half precision relative to float; despite doubling parallelism, the reduced DSP demand of half-precision arithmetic partially offsets the increased parallel factor. Mixed precision further slightly reduces DSP utilization by using fixed-point arithmetic for some operations.

\subsection{Model-Size Scaling}

We evaluated model-size scaling effects using the pneumonia dataset. The baseline inference-only kernel employs float precision with a hidden population configured as HCU=30, MCU=400, and sparsity parameters $n_\text{act}/n_\text{sil}$=320/80. The parallelism factor is fixed at eight.

\begin{figure}[htbp]
\centering
\begin{tikzpicture}
\begin{axis}[
    width=1\columnwidth,
    height=5cm,
    grid=major,
    xlabel={Latency (ms)},
    ylabel={Energy (mJ)},
    enlargelimits=true,
    scatter/classes={
        base={mark=diamond*,draw=black,fill=gray},
        hcu20={mark=triangle*,draw=black,fill=blue},
        hcu10={mark=triangle*,draw=black,fill=red},
        mcu300={mark=*,draw=black,fill=green},
        mcu200={mark=*,draw=black,fill=orange},
        nas160_40={mark=square*,draw=black,fill=purple},
        nas80_24={mark=square*,draw=black,fill=brown}
    },
    legend style={at={(0.5,-0.25)}, anchor=north, legend columns=3, font=\scriptsize}
]

\addplot[
    scatter,
    only marks,
    mark size=3.5pt,
    scatter src=explicit symbolic
]
coordinates {
    (18.759, 9.023) [base]           
    (11.869, 6.018) [hcu20]           
    (6.28, 3.096) [hcu10]           
    (13.519, 6.827) [mcu300]         
    (9.18, 4.618) [mcu200]          
    (10.705, 4.218) [nas160_40]       
    (6.831, 2.124) [nas80_24]        
};

\legend{
Base (86.2\%),
HCU = 20 (87.7\%),
HCU = 10 (84.3\%),
MCU = 300 (85.3\%),
MCU = 200 (86.5\%),
$n_\text{act}/n_\text{sil}$ = 160/40 (82.7\%),
$n_\text{act}/n_\text{sil}$ = 80/24 (64.1\%)
}

\end{axis}
\end{tikzpicture}
\caption{Latency vs Energy for model size variations on the Pneumonia dataset. The base model uses HCU=30, MCU=400, and $n_\text{act}/n_\text{sil}$=320/80. Accuracy is indicated in the legend.}
\label{fig:latency_energy_modelsize}
\end{figure}
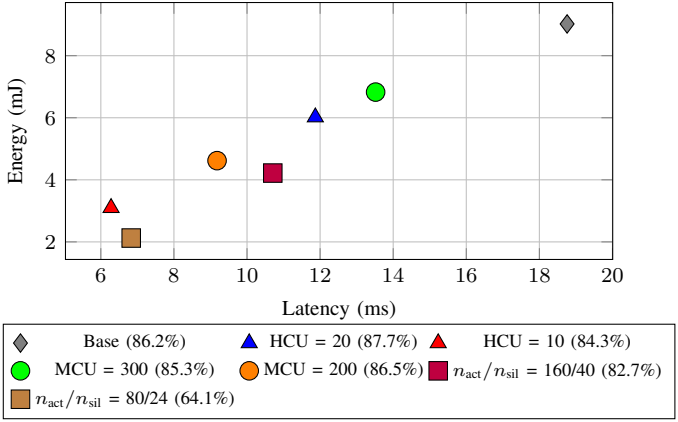

Figure~\ref{fig:latency_energy_modelsize} demonstrates how reducing model parameters affects latency, energy, and accuracy. Reductions in HCU strongly correlate with latency improvements, achieving up to 66\% latency reduction when HCU is decreased from 30 to 10. Similarly, sparsity of input-to-hidden connectivity ($n_\text{act}/n_\text{sil}$) notably reduces energy consumption. However, accuracy does not scale linearly: moderate parameter reductions (e.g., HCU=20, MCU=200) retain or slightly improve accuracy, whereas aggressive sparsification of connectivity ($n_\text{act}/n_\text{sil}$=80/24) significantly reduces accuracy to 64.1\%. Thus, connectivity sparsity parameters demand careful consideration due to their sensitivity in accuracy degradation.

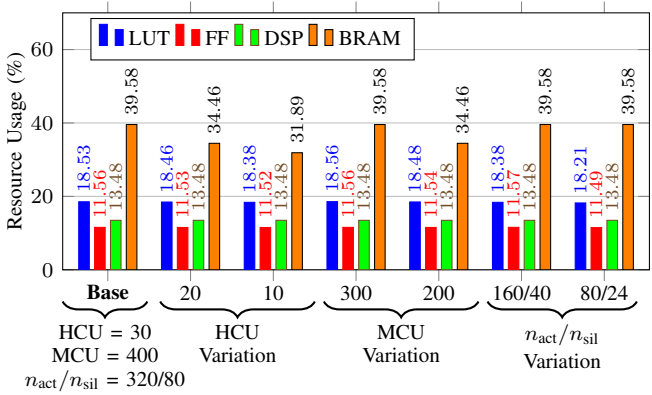
\begin{figure}[htbp]
\centering
\begin{adjustbox}{max width=\textwidth}
\begin{tikzpicture}
\begin{axis}[
    ybar,
    bar width=4pt,
    x=1.1cm, 
    width=20cm, height=5cm,
    enlarge x limits={abs=0.6cm},
    ylabel={\footnotesize Resource Usage (\%)},
    ylabel style={yshift=-5pt}, 
    symbolic x coords={base, hcu20, hcu10, mcu300, mcu200, rec5, rec25},
    xtick=data,
    xticklabels={\textbf{Base}, 20, 10, 300, 200, 160/40, 80/24},
    xticklabel style={align=center, yshift=2pt,rotate=0, font=\footnotesize},
    ymin=0,
    ymax=70,
    ymajorgrids=true,
    legend style={at={(0.5,-0.25)}, anchor=north, legend columns=4},
    nodes near coords,
    nodes near coords style={
        rotate=90,
        anchor=west,
        font=\scriptsize
    },
    legend style={
        at={(0.6,0.98)},
        anchor=north east,
        draw=black,
        fill=white,
        font=\footnotesize
    }
]

\addplot+[fill=blue]    coordinates {
(base,18.53) 
(hcu20,18.46) (hcu10,18.38) 
(mcu300,18.56) (mcu200,18.48) 
(rec5,18.38) (rec25,18.21)
};
\addplot+[fill=red]     coordinates {
(base,11.56) 
(hcu20,11.53) (hcu10,11.52) 
(mcu300,11.56) (mcu200,11.54) 
(rec5,11.57) (rec25,11.49)
};
\addplot+[fill=green]   coordinates {
(base,13.48) 
(hcu20,13.48) (hcu10,13.48) 
(mcu300,13.48) (mcu200,13.48) 
(rec5,13.48) (rec25,13.48)
};
\addplot+[fill=orange]  coordinates {
(base,39.58) 
(hcu20,34.46) (hcu10,31.89) 
(mcu300,39.58) (mcu200,34.46) 
(rec5,39.58) (rec25,39.58)
};
\legend{LUT, FF, DSP, BRAM}
\end{axis}

\draw[decorate,decoration={brace, mirror, amplitude=6pt}, thick, yshift=-12pt]
    ([xshift=0pt]0,0) -- ([xshift=32pt]0,0) node[midway, below=5pt, font=\footnotesize]{\shortstack{HCU = 30\\MCU = 400\\$n_\text{act}/n_\text{sil}$ = 320/80}};

\draw[decorate,decoration={brace, mirror, amplitude=6pt}, thick, yshift=-12pt]
    ([xshift=-20pt]2,0) -- ([xshift=10pt]3,0) node[midway, below=5pt, font=\footnotesize]{\shortstack{HCU\\Variation}};

\draw[decorate,decoration={brace, mirror, amplitude=6pt}, thick, yshift=-12pt]
    ([xshift=-15pt]4,0) -- ([xshift=15pt]5,0) node[midway, below=5pt, font=\footnotesize]{\shortstack{MCU\\Variation}};

\draw[decorate,decoration={brace, mirror, amplitude=6pt}, thick, yshift=-12pt]
    ([xshift=-10pt]6,0) -- ([xshift=18pt]7,0) node[midway, below=5pt, font=\footnotesize]{\shortstack{$n_\text{act}/n_\text{sil}$\\Variation}};

\end{tikzpicture}
\end{adjustbox}
\caption{FPGA Resource Usage for the pneumonia BLNN as model size scales}
\label{fig:precision_variant}
\end{figure}

Figure~\ref{fig:precision_variant} shows FPGA resource utilization across scaled models. Resource usage (LUT, FF, DSP, BRAM) remains nearly constant due to the dataflow architecture maintaining fixed parallelization and FIFO depths. The stable resource usage, despite significant performance variability, emphasizes the efficiency and flexibility of our architecture for scalable deployments. 

\subsection{Design Insights}

Based on extensive evaluations, we summarize key design insights and recommendations for deploying \gls{BLNN} accelerators on embedded FPGAs:

\paragraph{Kernel Choice and Resource Constraints}
The full kernel is feasible but inherently limited by BRAM, primarily due to storing multiple large synaptic traces required for synaptic plasticity updates. Therefore, its practical edge application is suitable only for smaller models.
The inference-only kernel significantly outperforms the full kernel in resource utilization and scalability, and is demonstrably more resource-efficient on embedded FPGAs. It supports larger models and higher-resolution inputs.

\paragraph{Precision Selection for Edge Deployment}
Half precision (FP16) provides optimal performance, offering significant latency and energy improvements with negligible accuracy degradation across tested datasets. Thus, FP16 precision is strongly recommended as a general-purpose configuration for inference-only edge \gls{BLNN} implementations.
Mixed precision (FXP16/FP16) shows moderate resource improvements but introduces non-trivial accuracy loss in more complex datasets, limiting its practical applicability. 

\paragraph{Recommendations on Model Scaling Priority}
To optimize latency and energy:
\begin{itemize}
    \item Prioritize reducing HCU first, as it significantly reduces latency with minimal accuracy impact.
    \item Then, scale down the MCU for additional latency and energy benefits without substantial accuracy loss.
    \item Lastly, carefully adjust connectivity sparsity parameters ($n_\text{act}/n_\text{sil}$), as aggressive sparsification can significantly compromise accuracy.
\end{itemize}


\section{Related Work}

Several prior efforts have explored FPGA-based implementations of \gls{BLNN} using the \gls{BCPNN} model \cite{liu_fpga-based_2020-1, podobas_streambrain_2021, wang_fpga-based_2024, wang_scalable_2025, al_hafiz_reconfigurable_2025}. Liu \textit{et al.} (2020) introduced a hardware accelerator for a spiking-based \gls{BCPNN} model on a low-cost Xilinx Artix-7 FPGA \cite{liu_fpga-based_2020}. Their work optimized computationally intensive synaptic state updates through a "lazy update mode," but comprehensive hardware support for \gls{BCPNN} training and inference remained unexplored \cite{liu_fpga-based_2020}. 

In 2021, Podobas \textit{et al.} proposed StreamBrain, a domain-specific language framework enabling \gls{BCPNN} deployment on \gls{HPC} systems using an Intel Stratix V FPGA \cite{podobas_streambrain_2021}. Their implementation demonstrated fast MNIST training and extended applicability to STL-10-sized networks, yet they limited FPGA acceleration to the two heaviest computational steps rather than a complete \gls{BCPNN} pipeline \cite{podobas_streambrain_2021}. 

Wang \textit{et al.} (2024) demonstrated a significant latency and power reduction in a \gls{BCPNN}-based associative memory system using a single Xilinx Alveo U200 FPGA compared to GPUs \cite{wang_fpga-based_2024}. They extended this work in 2025 with a scalable multi-FPGA architecture utilizing two Alveo U50 FPGA cards, showing further performance gains \cite{wang_scalable_2025}. However, their research primarily focused on associative memory architectures rather than evaluating \gls{BCPNN} using practical neural network datasets common in real-world applications.

More recently, Hafiz \textit{et al.} (2025) developed the first high-performance, reconfigurable, stream-based FPGA accelerator for rate-based \gls{BCPNN} implemented on a Xilinx Alveo U55C FPGA using Vitis \gls{HLS} \cite{al_hafiz_reconfigurable_2025}. Their approach emphasized optimizing computational units and synaptic plasticity for datacenter FPGAs equipped with \gls{HBM}. While powerful, this solution remains impractical for edge deployments, given its substantial power and resource demands \cite{al_hafiz_reconfigurable_2025}. 

Thus, existing FPGA-based \gls{BLNN} or \gls{BCPNN} implementations typically target \gls{HPC} or datacenter-grade hardware, leaving the domain of resource-constrained embedded edge FPGAs unexplored. Motivated by this gap, our work presents a \gls{BLNN} accelerator based on \gls{BCPNN} tailored specifically for embedded FPGA platforms. Our design uniquely supports a full online-learning kernel and a scalable inference-only kernel optimized for constrained hardware resources and stringent power budgets.
 
\section{Conclusion}

This work presented an FPGA accelerator for \glspl{BLNN} based on the \gls{BCPNN} model, on an embedded FPGA platform for both full online learning and inference-only kernels. We evaluated precision formats and model scalability under resource and performance constraints. 
The inference-only kernel proved significantly more efficient and scalable than the full kernel. Among tested formats, half precision (FP16) offered the best balance of latency, energy, and accuracy due to higher parallelism under limited memory bandwidth. Mixed precision showed accuracy loss on complex datasets with no clear advantage over FP16. Scaling experiments revealed that reducing hidden layer sizes (HCU/MCU) and increasing sparsity ($n_\text{act}/n_\text{sil}$) can lower latency and energy, though excessive sparsity degrades accuracy. Notably, the hardware utilization remained stable across scaling scenarios due to the stream-based accelerator architecture, making it well-suited for flexible deployment at the edge. Overall, the FP16 inference-only kernel stands out as a scalable, energy-efficient solution for embedded BLNNs, with future work aimed at broader application exploration.

\section*{Acknowledgment}

This work was funded by the European Commission Directorate-General for Communications Networks, Content and Technology, grant no. 101135809 (EXTRA-BRAIN), the Swedish Research Council grant no. 2021-04579 (Building Digital Brains), and the Swedish e-Science Research Centre (SeRC). The computations were enabled by resources provided by the Chalmers e-Commons at Chalmers and National Academic Infrastructure for Supercomputing in Sweden (NAISS), partially funded by the Swedish Research Council through grant agreement no. 2022-06725.

 \bibliographystyle{ieeetr}
 \bibliography{references,extra}

\end{document}